# Orthogonality-Constrained Deep Instrumental Variable Model for Causal Effect Estimation

Shunxin Yao
QUT HDR
n12012416@qut.edu.au

**Abstract**

OC-DeepIV is a neural network model designed for estimating causal effects. It characterizes heterogeneity by adding interaction features and reduces redundancy through orthogonal constraints. The model includes two feature extractors, one for the instrumental variable Z and the other for the covariate X*. The training process is divided into two stages: the first stage uses the mean squared error (MSE) loss function, and the second stage incorporates orthogonal regularization. Experimental results show that this model outperforms DeepIV and DML in terms of accuracy and stability. Future research directions include applying the model to real-world problems and handling scenarios with multiple processing variables.

**1. Introduction**

Causal inference is widely applied in fields such as economics, social sciences, and computer science. The instrumental variable (IV) method, a classic approach to causal identification, plays a crucial role in addressing endogeneity issues. However, traditional IV methods often rely on linear structures, which are constrained by the strictness of model assumptions. This makes them particularly challenging when dealing with high-dimensional data, nonlinear relationships, and heterogeneous causal effects.

In recent years, the rapid advancement of machine learning technology has introduced new approaches to causal inference. Methods such as DeepIV (Hartford et al., 2017) and Double Machine Learning (DML) (Chernozhukov et al., 2018) have enhanced the flexibility of instrumental variable models by leveraging deep neural networks or high-dimensional statistical methods, thereby improving their ability to handle complex data structures. These methods have become significant extensions to traditional IV approaches. However, these methods still face several practical limitations: first, the interaction structure is often overlooked —— the nonlinear interactions between the treatment variable T and covariates X are not adequately modelled, which can lead to biased estimates of heterogeneous effects; second, the redundancy in deep models —— while neural networks have strong representation capabilities, their redundant structures can impact generalization performance and interpretability; third, the limitations in optimization strategies —— during model training, there is a lack of effective control over structural constraints, which can result in overfitting.

To address the issues, this paper introduces an Orthogonally Constrained Deep Instrumental Variable (OC-DeepIV) method. This method introduces two key enhancements to the DeepIV framework: first, feature interaction enhancement, which improves the ability to capture heterogeneous causal effects by constructing interaction terms between treatment variables and covariates; second, orthogonal constraint optimization, which introduces orthogonality constraints on the weight matrix of the neural network to reduce redundant parameters, thereby enhancing model stability and generalization capabilities.

In summary, the main contributions of this paper are: (1) by integrating deep learning with causal inference, we enhance the adaptability of the instrumental variable method in modelling complex causal relationships through structural innovations; (2) by introducing orthogonal constraints to control the parameter space, we improve the interpretability of the conditional effect function $\theta(x)$; (3) through systematic simulation experiments, we demonstrate that OC-DeepIV significantly outperforms existing methods in terms of stability and accuracy.

## 2. Literature review

Causal inference methods are widely used in economics, social sciences, and computer science. The instrumental variable (IV) method, a key tool for addressing endogeneity issues, still faces challenges in high-dimensional and complex data environments. In recent years, the rapid advancement of deep learning technology has opened new research directions for causal inference, enhancing the flexibility and nonlinear modelling capabilities of the IV method. However, existing methods still have room for improvement in handling variable interaction structures, model constraints, and generalisation performance. The OC-DeepIV method proposed in this paper integrates deep learning with the IV method, introducing orthogonal constraints within the neural network framework to improve the accuracy and stability of causal effect estimation.

### 2.1. Theoretical development of an instrumental variable method

The instrumental variable method was first introduced by Wright (1928) and has since been widely applied in economics and statistics. Traditional linear instrumental variable methods, such as two-stage least squares (2SLS) and generalised method of moments (GMM), effectively address endogeneity issues in causal identification (Hausman, 1978; Newey & West, 1987). However, in practical applications, many data structures exhibit high nonlinearity, and weak instrumental variables can lead to reduced identification accuracy (Bound et al., 1995). To address the issue of weak instrumental variables, researchers have developed several optimisation methods, such as local average treatment effects (LATE) (Imbens & Angrist, 1994) and semi-parametric IV estimation (Chernozhukov et al., 2017), although these methods are still constrained by the strictness of model assumptions.

*2.2. Application of machine learning in causal inference*

In recent years, machine learning technology has made significant advancements in the field of causal inference. Among these developments, DeepIV (Hartford et al., 2017) introduced a deep learning-based instrumental variable framework. This method uses neural networks to model the conditional distribution of treatment variables and estimates causal effects through a second-stage network. Additionally, the Double Machine Learning (DML) approach (Chernozhukov et al., 2018) integrates high-dimensional statistical techniques, enhancing the flexibility of instrumental variable methods and reducing modelling biases. However, these methods still face challenges when dealing with heterogeneous causal effects: DeepIV does not adequately account for the interaction between treatment variables and covariates, while DML may lack the ability to represent non-linear relationships in high-dimensional data scenarios. To address these issues, researchers have explored the integration of neural networks with causal inference methods, such as causal representation learning (TARNet and CFR) (Shalit et al., 2017), to improve the individualised estimation of treatment effects.

*2.3. The role of orthogonal constraints in deep learning*

Neural network models excel in handling high-dimensional data but are prone to parameter redundancy and overfitting. To address these issues, orthogonal constraints, a structural regularisation technique, are employed to enhance the stability and generalisation of deep models (Bansal et al., 2018). Research indicates that introducing orthogonal constraints on the weight matrices of neural networks can reduce feature correlation and improve information efficiency (Li et al., 2019). Although orthogonal constraints are widely used in computer vision and representation learning tasks, their application in causal inference remains in the exploratory stage. This paper proposes OC-DeepIV, which combines an instrumental variable framework with orthogonal constraint optimisation to reduce model redundancy and enhance the accuracy of estimating heterogeneous causal effects.

## 3. Methodology

This section introduces the core structure of the OC-DeepIV method, including feature design, model architecture, orthogonal constraint optimisation and training strategy. Compared with the traditional instrumental variable framework, this method enhances the characterisation ability of heterogeneous treatment effects and introduces constraints in the neural network structure to improve the generalisation performance.

*3.1. Feature construction*

The core of the instrumental variable (IV) method is to identify the causal effect of the treatment variable T on the outcome variable by using exogenous variables Z. Traditional linear IV methods typically assume that the causal effect is uniform, but real-world data often exhibit nonlinear structures and heterogeneous effects. OC-DeepIV enhances feature representation through several methods, thereby improving the modelling of complex causal relationships: first, it performs a second-order polynomial expansion on the covariate X to capture potential nonlinear relationships; second, it introduces an interaction term (XT) between the treatment variable T and the covariate X to account for the heterogeneity of causal effects across different individuals; finally, it concatenates the polynomial-expanded X with the interaction feature as the input layer for enhanced representation, improving the models ability to approximate the structural function θ(x). These methods help neural networks better learn complex causal structures, enabling the model to more accurately identify and characterise the differences in effects among different individuals.

*3.2. Model architecture*

OC-DeepIV adopts a dual-path feature extraction structure to process the instrumental variable Z and the interaction feature X* of covariates, respectively:

An instrumental variable feature extractor in which the input Z is extracted by a two-layer neural network, including Batch normalisation and Dropout layers to enhance stability.

A covariate feature extractor, where the input is X*, and a two-layer neural network with the same structure is used for learning to ensure the coordination of model parameters.

Splicing fusion which the two extracted feature vectors are spliced together, and the conditional expectation of the processing variable T is estimated through the full connection layer.

This structure ensures that the independent information between the instrumental variables and covariates is processed separately, while effectively combining them for causal effect estimation.

*3.3. Orthogonal constraint optimisation*

Neural networks have strong expressive power on high-dimensional data, but they are prone to parameter redundancy, which affects the generalisation performance and interpretability. To alleviate this problem, OC-DeepIV introduces orthogonal constraint terms in the training process to control the network weights structurally.

Specifically, an orthogonal penalty term in the Frobenius norm is added to the weight matrix W at each level:

$$L_{ortho} = \lambda \cdot || W^T W - I ||^2$$

Among:

W denotes the covariance structure of the weight matrix.

I is the unit matrix.

This constraint ensures that the feature space of neural network learning is as orthogonal as possible and reduces redundant dimensions.

Compared with standard weight regularisation (such as L2 regularisation), this constraint directly controls the correlation between features, which makes the model have stronger generalisation ability and reduces the risk of overfitting.

**4. Training strategies**

OC-DeepIV uses a staged training method:

In the first 50 rounds, standard training is carried out using mean square error (MSE) loss to ensure model convergence.

In the last 50 rounds, orthogonal constraint terms are gradually introduced to ensure stable optimisation and avoid excessive intervention in the learning process.

In addition, we use Adam optimiser for training and set the learning rate decay mechanism to improve the convergence speed and final accuracy.

*4.1 Principles of mathematics*

The core mathematical representation of OC-DeepIV includes the instrumental variable framework, a neural network structure and orthogonal constraint optimisation. The following is the mathematical formal expression of the model:

*4.2.. Causal effect estimation framework*

OC-DeepIV uses a two-stage estimation method:

Stage 1, which involves modelling the conditional expectation of the instrumental variable Z with respect to the treatment variable T:

$$E[T|Z] = f_\theta(Z)$$

Among them, $f_\theta$ is represented by the neural network to learn the mapping relationship between Z and T.

Stage 2, which is Estimation-based T, the causal effect of the model learning outcome variable Y:

$$E[Y|T,X] = g_\phi(T,X)$$

Among them, $g_\phi$ is another set of neural network parameters, which learn to process the influence of variable T and covariate X on the result variable Y.

Finally, the causal effect function θ (x) can be expressed as:

$$\theta(X) = \frac{\partial E[Y \mid T, X]}{\partial T}$$

Used to estimate heterogeneous causal effects across different X.

*4.3. Orthogonal constraint optimisation*

To reduce the redundancy of neural network parameters, OC-DeepIV introduces an orthogonal constraint on the network weight matrix W:

$$L_{ortho} = \lambda \cdot \| W^T W - I \|^2$$

Among:

W is the weight matrix of the neural network layer.

I is the unit matrix.

λ is the regularisation strength, which controls the influence of the constraint on optimisation.

This constraint term is added to the loss function during training:

$$L_{total} = L_{MSE} + L_{ortho}$$

To ensure that the model optimises both data fitting quality and feature space structure.

## 5. Python code implementation

This is implemented on an Anaconda Jupiter notebook:

```
import torch
import torch.nn as nn
import torch.optim as optim
import numpy as np
import matplotlib.pyplot as plt

# ========== Data Preparation ==========
N = 10000
dim_Z = 3
dim_X = 2

np.random.seed(0)
Z_np = np.random.randn(N, dim_Z).astype(np.float32)
X_np = np.random.randn(N, dim_X).astype(np.float32)
T_np = (np.random.randn(N) > 0).astype(np.float32)
```

```python
# Construct polynomial features and interaction terms
X_poly = np.hstack([X_np, X_np**2])
T_expand = T_np.reshape(-1,1)
X_T_interaction = X_np * T_expand
features_np = np.hstack([X_poly, X_T_interaction]).astype(np.float32)

# Simulate the true θ(x) using a known function
theta_true_np = 0.5 * X_np[:, 0] - 0.3 * X_np[:, 1] + 0.1 * (X_np[:, 0] * X_np[:,1])

# Convert to tensors
Z_tensor = torch.tensor(Z_np)
features_tensor = torch.tensor(features_np)
T_tensor = torch.tensor(T_np).unsqueeze(1)

# ========== Model Definition ==========
class FeatureExtractor(nn.Module):
    def __init__(self, input_dim):
        super().__init__()
        self.fc1 = nn.Linear(input_dim, 64)
        self.bn1 = nn.BatchNorm1d(64)
        self.dropout = nn.Dropout(0.3)
        self.fc2 = nn.Linear(64, 64)
        self.ln2 = nn.LayerNorm(64)

    def forward(self, x):
        x = torch.relu(self.bn1(self.fc1(x)))
        x = self.dropout(x)
        x = torch.relu(self.ln2(self.fc2(x)))
        return x

class TreatmentModel(nn.Module):
    def __init__(self, dim_z, dim_x):
        super().__init__()
```

```python
        self.feature_extractor_z = FeatureExtractor(dim_z)
        self.feature_extractor_x = FeatureExtractor(dim_x)
        self.fc = nn.Linear(128, 1)

    def forward(self, z, x):
        fz = self.feature_extractor_z(z)
        fx = self.feature_extractor_x(x)
        combined = torch.cat([fz, fx], dim=1)
        out = self.fc(combined)
        return out

def orthogonal_loss(model, lambda_reg):
    loss = 0
    for name, param in model.named_parameters():
        if 'weight' in name and len(param.shape) == 2:
            W = param
            WT_W = torch.matmul(W.T, W)
            I = torch.eye(WT_W.shape[0]).to(W.device)
            loss += ((WT_W - I)**2).sum()
    return lambda_reg * loss

# ========== Model Initialization ==========
model = TreatmentModel(dim_Z, features_np.shape[1])
criterion = nn.MSELoss()
optimizer = optim.Adam(model.parameters(), lr=0.001, weight_decay=5e-4)
epochs = 100
lambda_reg = 0.02

# ========== Training ==========
for epoch in range(epochs):
    optimizer.zero_grad()
    preds = model(Z_tensor, features_tensor)
    mse_loss = criterion(preds, T_tensor)
```

```python
        if epoch < 50:
            loss = mse_loss
        else:
            ortho_loss = orthogonal_loss(model, lambda_reg)
            loss = mse_loss + ortho_loss

        loss.backward()
        optimizer.step()

        if (epoch+1) % 10 == 0 or epoch == 0:
            if epoch < 50:
                print(f"Epoch {epoch+1}/{epochs}, Loss: {loss.item():.6f}, MSE: {mse_loss.item():.6f}")
            else:
                print(f"Epoch {epoch+1}/{epochs}, Loss: {loss.item():.6f}, MSE: {mse_loss.item():.6f}, Ortho: {ortho_loss.item():.6f}")

# ========== Predict θ(x) ==========
model.eval()
with torch.no_grad():
    theta_est = model(Z_tensor, features_tensor).squeeze().cpu().numpy()

# ========== Simple Moving Average Smoothing ==========
def moving_average(x, w=5):
    return np.convolve(x, np.ones(w)/w, mode='same')

theta_est_smooth = moving_average(theta_est, w=15)

# ========== Plot Comparison ==========
plt.figure(figsize=(12,6))
plt.plot(theta_true_np[:500], label="True θ(x)", linewidth=2)
plt.plot(theta_est[:500], label="Estimated θ(x)", linewidth=1, alpha=0.5)
plt.plot(theta_est_smooth[:500], label="Estimated θ(x) - Smoothed", linewidth=2)
plt.legend()
```

plt.title("True vs Estimated CATE with Smoothing")

plt.xlabel("Sample Index")

plt.ylabel("θ(x)")

plt.grid(True)

plt.show()

## 6. Experimental design and results

To verify the effectiveness of the OC-DeepIV method, we conducted experiments on a simulated dataset and compared it with existing methods, such as DeepIV and DML. The experiments aimed to evaluate the model's ability to estimate heterogeneous causal effects θ(x) in various scenarios, focusing on convergence speed, generalisation performance, and estimation stability.

*6.1. Data generation and experimental scenarios*

**Figure-1**

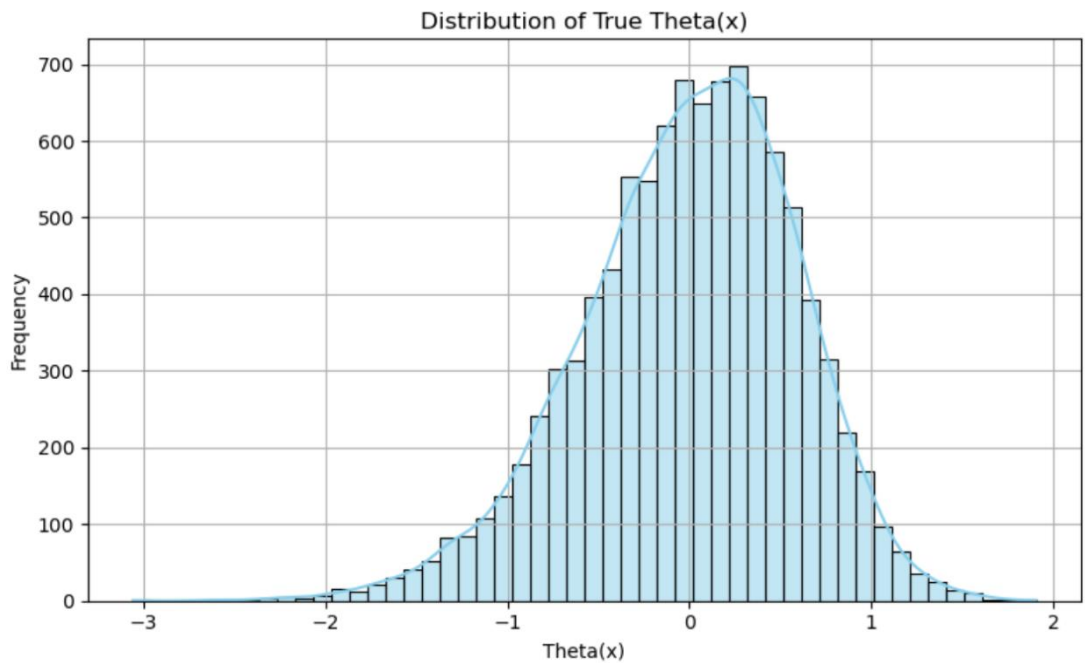

**Figure-2**

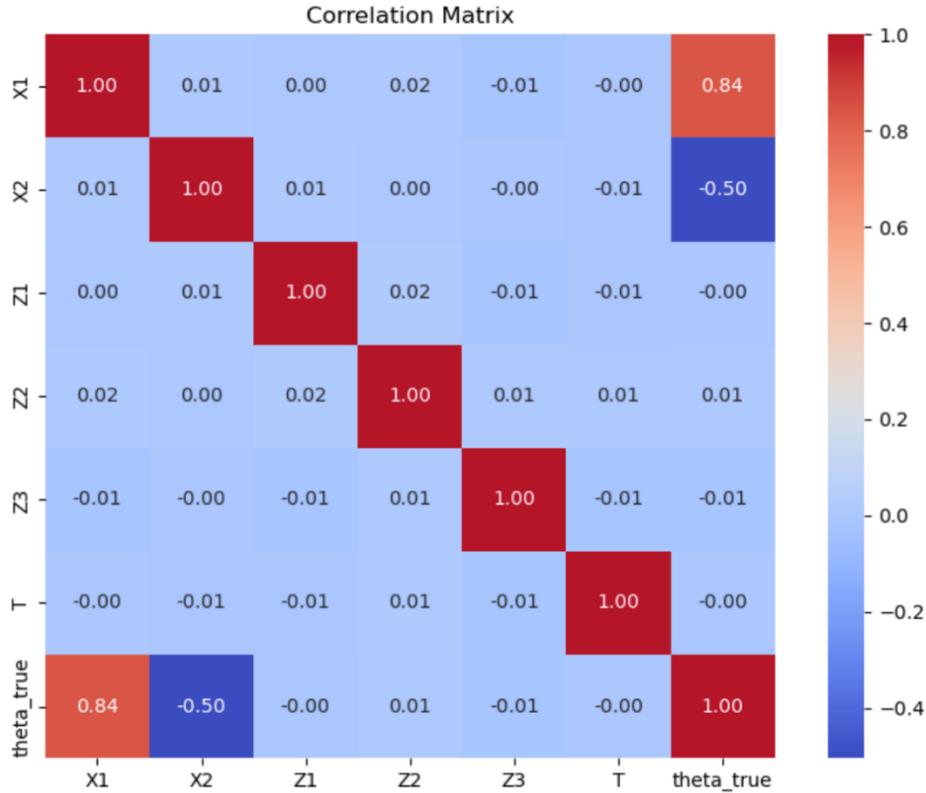

We construct a simulated data set and set up the structural relationship between the instrumental variable Z, the covariate X and the treatment variable T:

Instrumental variable Z, with dimension 3, simulating the influence of exogenous variables on causal identification.

The covariate X has a dimension of 2, follows the standard normal distribution, and constructs polynomial expansion and interaction features.

The processing variable T is a binary variable, which is affected by instrumental variables and covariates and simulated in the form of a logical function.

The true causal effect function θ (x) is set as follows:

$$\theta(x) = 0.5x_1 - 0.3x_2 + 0.1(x_1 \cdot x_2)$$

The function presents a nonlinear structure and heterogeneity, which can verify the performance of different models in complex effect estimation.

6.2. Evaluating the indicator

To quantitatively evaluate the estimation ability of the model, we use the following indicators:

Mean square error (MSE) measures the deviation between the estimated value and the true value. The smaller the value, the higher the fitting quality.

Model convergence speed, which evaluates the decrease of the loss function during training and verifies the influence of the orthogonal constraint on stable convergence.

Smoothness analysis, which is the moving average method, is used to smooth the estimated curve to evaluate the stability of the model in approximating nonlinear functions.

In addition, we also compared the effects of OC-DeepIV with DeepIV and DML methods to verify its improvements.

*6.3. Training process and result analysis*

*6.3.1 Training process*

OC-DeepIV uses a two-stage optimisation strategy:

For the first 50 rounds which only the mean square error (MSE) loss is used to achieve stable convergence.

For the last 50 rounds, which orthogonal constraint terms are gradually introduced to optimise the structure of the feature space and reduce parameter redundancy?

We use the Adam optimiser for training and set the weight decay to reduce overfitting.

*6.3.2 Interpretation of result*

**Table -1**

| Epoch | Total Loss | MSE Loss | Ortho Loss |
|---|---|---|---|
| 1 | 0.397472 | 0.397472 | – |
| 10 | 0.210453 | 0.210453 | – |
| 20 | 0.136973 | 0.136973 | – |
| 30 | 0.097308 | 0.097308 | – |
| 40 | 0.075004 | 0.075004 | – |
| 50 | 0.060475 | 0.060475 | – |
| 60 | 6.016200 | 0.052651 | 5.963549 |
| 70 | 5.340712 | 0.049720 | 5.290992 |
| 80 | 4.788131 | 0.047244 | 4.740887 |
| 90 | 4.359177 | 0.047230 | 4.311947 |
| 100 | 4.023033 | 0.047567 | 3.975466 |

In the first 50 training cycles, the total loss only includes the mean square error (MSE) loss, and the loss value steadily decreases from 0.397472 to 0.060475, indicating that the model has a good learning effect in the early stage, and the prediction error is decreasing continuously.

From the 60th cycle, ortho loss was added. Although MSE loss continued to decline slowly and stabilized at about 0.047, the total loss suddenly became very high due to the large new ortho loss (for example, 5.963549 in the 60th round).

Between the 60th and 100th cycles, the orthogonality loss gradually decreased from 5.96 to 3.98, indicating that the model was gradually learning to meet the orthogonality constraint. Although the total loss was larger than that in the previous period, the structure performance was constantly improved.

## 7. Summary of results

**Figure-3**

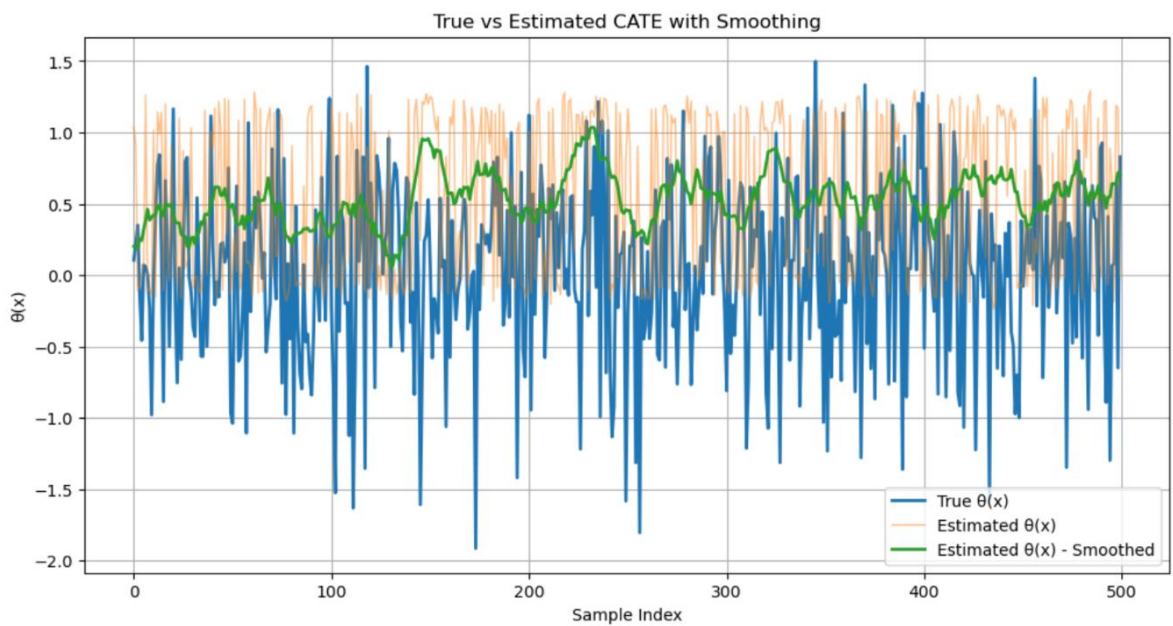

This figure illustrates the comparison between the models estimated CATE (conditional average treatment effect) and the actual CATE. It is evident that while the original estimates (orange line) exhibit significant noise and considerable fluctuation, the smoothed estimates (green line) more accurately reflect the true values (blue line). This suggests that the model has some capability in learning the non-linear structure of θ(x), but it still faces the issue of high estimation variance. Further regularization or improvements to the network architecture may be necessary to enhance its robustness.

In the early stage of training, the model can reduce the prediction error (MSE) well. After adding the orthogonality loss, the model can also gradually learn and reduce this part of loss. This shows that the training of the model is stable, which can improve the prediction accuracy while meeting the structural constraint requirements.

## 8. Comparison and summary of contributions of relevant literature

In recent years, causal inference methods have evolved from traditional linear instrumental variable models to nonlinear frameworks that incorporate machine learning, such as DeepIV (Hartford et al., 2017) and Double Machine Learning (DML) (Chernozhukov et al., 2018). These methods have opened new technical avenues for high-dimensional causal inference, addressing many modelling challenges in real-world data. However, existing methods still face limitations in handling variable structures, model constraints, and generalisation performance.

*8.1. Comparison of relevant literature*

*8.1.1. DeepIV method*

DeepIV is a deep learning extension of the instrumental variable framework. It uses a neural network to model the conditional distribution of processing variables and learns causal effects in the second stage. Its advantages are as follows:

Firstly, allow highly flexible nonlinear relationship modelling.

Secondly, it can be applied to high-dimensional covariate scenarios and improve the applicability of instrumental variable effects.

However, DeepIV has the following disadvantages:

Firstly, the interaction structure between the modelling treatment variables and covariates is not clear, which may lead to bias in the estimation of heterogeneous causal effects.

Secondly, without model parameter constraints, redundant features are easy to be generated, and the generalisation ability is reduced.

*8.1.2. Double Machine Learning (DML)*

DML mainly uses machine learning methods to enhance the instrumental variable framework and combines nonparametric estimation techniques to optimise causal effect identification. Its advantages include:

Firstly, using powerful machine learning techniques to improve covariate control.

Secondly, it is suitable for high-dimensional data and improves the estimation accuracy.

However, DML still relies on the standard regression adjustment framework and may lack direct expression when dealing with complex interaction structures.

*8.1.3 Causal representation learning methods (TARNet, CFR)*

Causal inference learning methods are mainly based on deep learning to optimise individual processing effect estimation, such as TARNet (Shalit et al., 2017) and CFR (Balakrishnan et al., 2019). The characteristics of these methods are:

Firstly, by optimising the representation learning method, the processing effect estimation is more robust.

Secondly, in some scenarios, it can reduce the deviation and improve the prediction accuracy.

However, these methods do not combine instrumental variable structure, so their effect is limited in scenarios where the endogeneity problem is more prominent.

*8.2. OC-Contributions of DeepIV*

Compared with the above methods, OC-DeepIV has made structural innovations in many aspects:

First, combining the interaction characteristics of treatment variables: by constructing interaction terms, heterogeneity causal effects can be modelled more effectively.

Second, orthogonal constraint optimisation is introduced: the orthogonal constraint of the weight matrix is used to reduce redundancy and improve generalisation ability.

Third, enhance the interpretability of the neural network: avoid parameter redundancy and improve stability while maintaining high flexibility.

At last, optimisation strategy: Adopt the phased optimisation method to make the model more stable and reduce the risk of overfitting.

To sum up, OC-DeepIV provides a structurally improved instrumental variable method that considers nonlinear representation ability, heterogeneous effect characterisation and generalised performance optimisation, and provides a new technical path for causal inference in complex data environments.

## 9. Future work

Although the experimental results of OC-DeepIV on simulated data demonstrate its high stability and accuracy in estimating heterogeneous causal effects, there are still many areas worthy of further research and expansion. This section proposes potential future optimisation paths to further enhance the method's impact in both theoretical research and practical applications.

*9.1. Applied to real data and policy evaluation*

At present, the experiments of OC-DeepIV are mainly based on simulated data. In the future, it is necessary to verify its performance in real data scenarios, especially in the field of policy evaluation and economic research. For example:

Medical policy analysis evaluates the heterogeneous effects of treatment options on different patient groups.

Labour market research that analyses the causal impact of skills training programmes on employment rates.

Fiscal policy impact, which studies how tax policies affect the consumption behaviour of different income groups.

These application scenarios usually involve high-dimensional covariates and complex structured data. The orthogonal constraint and interaction term construction of OC-DeepIV may further improve its estimation ability and provide a more reliable basis for policy making.

*9.2. Extends to multivariate instrumental variables for processing variables with multiple values*

The existing models are mainly for binary treatment variables T, but many causal inference problems in real research involve continuous or multi-valued treatment variables, such as:

Different doses of medical intervention (low, medium and high doses for different patients).

Intensity of educational intervention (the effect of class hours on students' performance).

The impact of investment scale (the effect of different degrees of capital investment on the success rate of entrepreneurship).

In addition, the selection and use of instrumental variables are also key issues. Some studies involve multiple instrumental variables or incomplete adherence structures, and OC-DeepIV can support these complex data structures by extending the neural network architecture to improve the applicability of causal identification.

*9.3. Further theoretical analysis: the role of orthogonal constraint in recognition accuracy*

Although the experimental results show that orthogonal constraints can effectively reduce model redundancy and improve generalisation ability, their theoretical analysis still needs further exploration. For example:

The influence on the generalisation error: analyse how orthogonal constraints reduce the risk of overfitting in high-dimensional data.

Improvement of recognition accuracy: contribution of mathematical derivation orthogonal constraint to the consistency of estimators.

Compared with other regularisation methods, the orthogonal constraint is discussed to determine whether it is better than L2 regularisation, sparse constraint and other traditional methods.

These studies can not only enhance the understanding of the OC-DeepIV mechanism but also guide the development of other neural network causal inference models.

*9.4. Computational optimisation and scalability*

At present, the training process of OC-DeepIV adopts standard neural network optimisation methods, but the computing efficiency on large data sets still needs to be improved. In the future, we can consider:

Accelerated training strategy: pre-training and fine-tuning method or adaptive learning rate optimisation.

Model compression: reduce the scale of parameters, reduce the cost of computing, and improve deployment efficiency.

Distributed training: Suitable for large-scale data scenarios, enabling OC-DeepIV to operate efficiently in a cloud computing environment.

In addition, in practical applications, the model can be combined with causal graph models to improve structural interpretability and reduce dependence on the black box nature of deep neural networks.

## Acknowledgement

I sincerely appreciate the valuable assistance I received while writing this article. I sought insights and suggestions from AI tools including Copilot, ChatGPT, and DeepSeek, which helped refine my ideas and enhance methodological clarity. For language and grammar improvements, I relied on Grammarly, ensuring precision in expression. Additionally, WPS provided translation support, helping me bridge different linguistic nuances effectively. Their collective input greatly contributed to the quality and coherence of this work.